\documentclass[superscriptaddress,altaffilletter,showkeys,endfloats*,preprint,longbibliography]{revtex4-1}
\usepackage{graphicx,amsmath,natbib}
\RequirePackage{lineno}
\begin{document}

\title{Hybrid Metal-Graphene Plasmons for Tunable Terahertz Technology}

\author{Mohammad M. Jadidi}
\email{mmjadidi@umd.edu}
\affiliation{Institute for Research in Electronics \& Applied Physics, University of Maryland, College Park, MD 20742, USA}

\author{Andrei B. Sushkov}
\affiliation{Center for Nanophysics and Advanced Materials, University of Maryland, College Park, Maryland 20742, USA}

\author{Rachael L. Myers-Ward}
\affiliation{US Naval Research Laboratory, Washington DC 20375, USA}

\author{Anthony K. Boyd}
\affiliation{US Naval Research Laboratory, Washington DC 20375, USA}

\author{Kevin M. Daniels}
\affiliation{US Naval Research Laboratory, Washington DC 20375, USA}

\author{D. Kurt Gaskill}
\email{kurt.gaskill@nrl.navy.mil }
\affiliation{US Naval Research Laboratory, Washington DC 20375, USA}

\author{Michael S. Fuhrer}
\email{michael.fuhrer@monash.edu}
\affiliation{Center for Nanophysics and Advanced Materials, University of Maryland, College Park, Maryland 20742, USA}
\affiliation{School of Physics, Monash University, 3800 Victoria, Australia}

\author{H. Dennis Drew}
\email{hdrew@umd.edu}
\affiliation{Center for Nanophysics and Advanced Materials, University of Maryland, College Park, Maryland 20742, USA}

\author{Thomas E. Murphy}
\email{tem@umd.edu}
\affiliation{Institute for Research in Electronics \& Applied Physics, University of Maryland, College Park, MD 20742, USA}

\begin{abstract}

Among its many outstanding properties, graphene supports terahertz surface plasma waves -- sub-wavelength charge density oscillations connected with electromagnetic fields that are tightly localized near the surface\cite{Fei2012,Chen2012}.  When these waves are confined to finite-sized graphene, plasmon resonances emerge that are characterized by alternating charge accumulation at the opposing edges of the graphene.  The resonant frequency of such a structure depends on both the size and the surface charge density, and can be electrically tuned throughout the terahertz range by applying a gate voltage\cite{Ju2011,Yan2012}.  The promise of tunable graphene THz plasmonics has yet to be fulfilled, however, because most proposed optoelectronic devices including detectors, filters, and modulators \cite{Hartmann2014,Vicarelli2012,Cai2014,Xing2012,Lee2012,Shi2014} desire near total modulation of the absorption or transmission, and require electrical contacts to the graphene -- constraints that are difficult to meet using existing plasmonic structures.  We report here a new class of plasmon resonance that occurs in a hybrid graphene-metal structure.  The sub-wavelength metal contacts form a capacitive grid for accumulating charge, while the narrow interleaved graphene channels, to first order, serves as a tunable inductive medium, thereby forming a structure that is resonantly-matched to an incident terahertz wave.  We experimentally demonstrate resonant absorption near the theoretical maximum in readily-available, large-area graphene, ideal for THz detectors and tunable absorbers.  We further predict that the use of high mobility graphene will allow resonant THz transmission near 100\%, realizing a tunable THz filter or modulator.  The structure is strongly coupled to incident THz radiation, and solves a fundamental problem of how to incorporate a tunable plasmonic channel into a device with electrical contacts.

\end{abstract}%


\maketitle
In order to be applied in practical optoelectronic devices, graphene terahertz plasmonic resonators must be connected to an antenna, transmission line, metamaterial, or other electrical contact, in order to sense or apply a voltage or current, or to improve the coupling to free-space radiation. The conductive boundary screens the electric field and inhibits the accumulation of charge density at the opposing edges of the graphene channel, thus disrupting the traditional graphene plasmon mode.  Until now, there was no experimental evidence that two-dimensional plasmons could be confined with conductive boundaries.

In this letter, we demonstrate a new type of plasmon resonance in metal-contacted graphene, and we use analytic calculations, numerical simulations, and THz reflection and transmission measurements to confirm the principle of operation.  These plasmon modes shows strong coupling to incident terahertz radiation, so that maximal absorption in graphene can be achieved at a resonance frequency that is gate-tunable. We also introduce an equivalent circuit model that predicts the resonant frequency, linewidth, and impedance matching condition of the fundamental plasmon mode, and can be used for designing graphene plasmonic metamaterials and antenna coupled devices.  We present predicted results for high mobility graphene that show that a near 100\% tunable resonant transmission can be achieved, giving an ideal platform for THz modulators and tunable bandpass filters.

\begin{figure}
  \centering
  \includegraphics[scale=1.5]{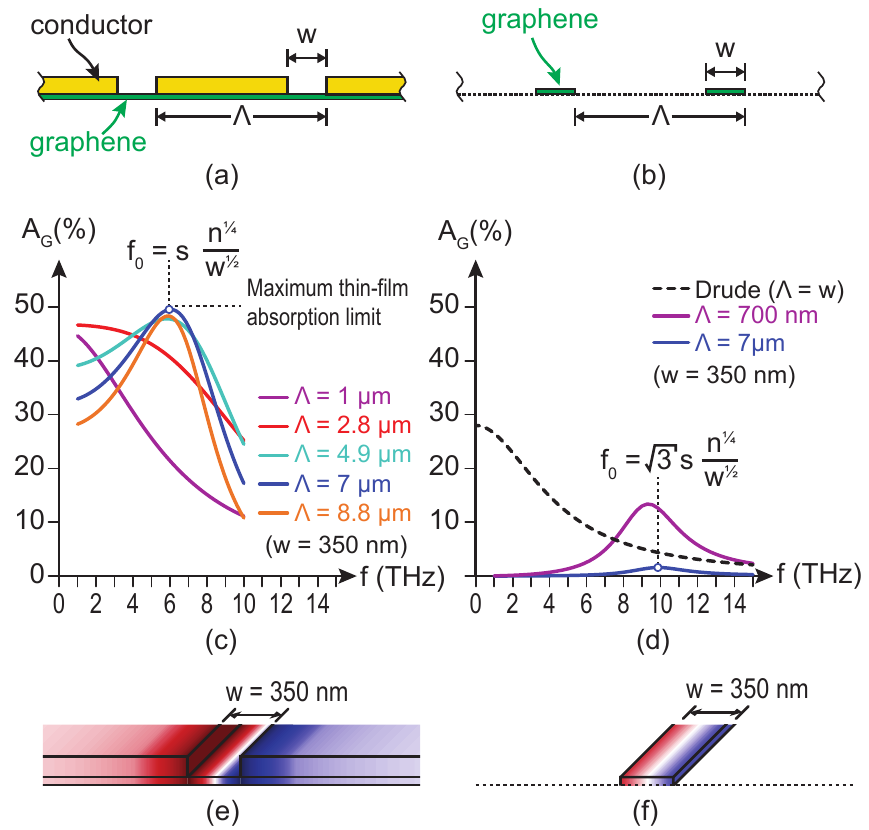}
  \caption{(a) Geometry of hybrid metal-graphene structure considered here. (b) Comparable array of isolated graphene ribbons.  (c) Calculated graphene absorption spectrum $A_G(\omega)$ for the hybrid metal-graphene device with periods of $\Lambda =$ 1, 2.8, 4.9, 7 and 8.8 $\mu$m, for a graphene channel with $w= 350$ nm, $n=1.5 \times 10^{13}$ cm$^{-2}$, and  $\mu=1000$ cm$^{2}V^{-1}s^{-1}$.  The upper and lower dielectric regions were assumed to be identical, with $\epsilon_1=\epsilon_2=5$, in which case the theoretical maximum thin-film absorption is 50\%\cite{Hilsum1954}, indicated by the horizontal dashed line. (d) Calculated absorption spectrum for isolated graphene ribbons with material properties identical to the channels considered in (c), and periods of $\Lambda =$ 0.7 and 7 $\mu$m.  For comparison, the dashed line indicates the Drude absorption spectrum for a continuous graphene sheet. (e)/(f) Calculated charge density profile at the plasmon resonant frequency for the hybrid metal-graphene device and graphene ribbon, respectively.  }
  \label{fig:1}
\end{figure}

Fig.~\ref{fig:1}a shows the structure of the metal-contacted graphene plasmonic device considered here, which is comprised of a periodic array of narrow slots in a metallic layer that is patterned on top of a continuous graphene layer. For comparison, in Fig.~\ref{fig:1}b we also consider an array of isolated graphene ribbons of comparable dimension.  In both cases, the period $\Lambda$ is taken to be small compared to the free-space wavelength.  To calculate the plasmon resonances and absorption in these structures, we adapt the method of \cite{Mikhailov2006} to obtain an integral equation for the in-plane electric field when the structure is illuminated by a normally incident plane wave at frequency $\omega$ that is linearly polarized in the direction perpendicular to the graphene channels.  The theory is presented in the supplementary section (SS1).  The resonant modes and fractional absorption in the graphene $A(\omega)$ is then found by integrating the Joule power density over the graphene ribbon, and normalizing to the incident power of the plane wave (S5).  The calculated absorption spectrum reveals all of the dipole-active plasmon resonances and the relative coupling of these modes to radiation.  

In Fig.~\ref{fig:1}c we present the theoretically computed absorption spectrum $A(\omega)$ for several different metal periods $\Lambda$, with the graphene ribbon width $w = 350$ nm held constant.  The mobility and carrier density (electron or hole) were taken to be $\mu=1000$ cm$^2$V$^{-1}$s$^{-1}$ and $n=1.5\times10^{13} $ cm$^{-2}$, respectively. The array shows no discernable plasmon resonance when the period $\Lambda$ and graphene width $w$ are comparable, giving instead a Drude-like response. However, when the metal contacts are made much wider than the graphene channel, a strong resonance emerges, characterized by high absorption in the graphene ribbon, at a resonant frequency that scales with $n^{1/4}w^{-1/2}$, similar to the plasmon resonances in uncontacted graphene ribbons \cite{Ju2011,Nene2014}. The surrounding material is assumed to be a uniform dielectric, in which case, the maximum achievable absorption in a thin-film (metal-graphene grating) is 50\% \cite{Hilsum1954} (also S26 in Supplementary Section). As shown in Fig.~\ref{fig:1}c, at the resonant frequency, the graphene absorption reaches a peak of the maximum possible value (50\%), even when the geometrical fill factor is only $w/\Lambda = 1/20$ (5\%). This suggests an extremely high confinement of the THz field in the narrow slots where graphene is located. We note that by using known techniques such as an anti-reflection coating or a Salisbury screen \cite{Jang2014} on top of the grating, the thin film limit absorption can be increased to nearly 100\%, and a perfect tunable graphene plasmonic  absorber can be achieved. The calculations confirm that these resonances disappear when the graphene is absent, when the polarization is rotated parallel to the channels, or when the graphene is electrostatically gated to the charge neutral point.  For comparison, in Fig.~\ref{fig:1}d, we show the absorption spectrum for an array of electrically isolated graphene ribbons of identical width, carrier density, and mobility, which yields a far lower on-resonant absorption (blue curve), even when the fill-factor is increased to 50\% (purple curve).

The nature of the fundamental metal-graphene plasmon resonance is illustrated in Fig.~\ref{fig:1}e, which shows the charge density calculated at the resonant frequency.  For comparison, we also show in Fig.~\ref{fig:1}f the familiar plasmon resonance for an uncontacted graphene ribbon of the same dimension. In the contacted graphene, the metal regions act as capacitive reservoir for charge accumulation, and the graphene serves as an inductive channel, thus forming a resonant circuit that interacts strongly with the incident radiation.  This is in striking contrast to the isolated ribbon case, where the coupling to incident radiation is weaker, and does not depend sensitively on the grating period\cite{Ju2011,Yan2012,Nene2014}. The extension of the spatial mode also explains the significant reduction of plasmon frequency (predicted by the theory in Supplementary Section SS1 for the case $\Lambda \gg w$) which is reduced by about a factor of $\sqrt 3$ compared to that of an isolated graphene ribbon \cite{Mikhailov2006}.

\begin{figure}
  \centering
  \includegraphics[scale=1.5]{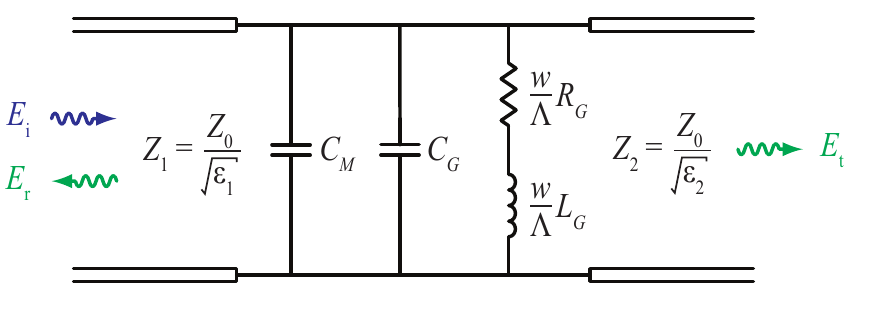}
  \caption{Two-port equivalent circuit used to model for the hybrid metal-graphene grating. $R_G$ and $L_G$ are the graphene ohmic resistance and kinetic inductance respectively. $C_G$ is the graphene ribbon array capacitance, and $C_M$ is the capacitance of the metallic grid. The transmission ($T=\sqrt{\epsilon_2/\epsilon_1}|E_t/E_i|^2$), reflection($R=|E_r/E_i|^2$), and graphene absorption ($1-R-T$) can be approximately found from this circuit (S14,S15).  $Z_1 \equiv Z_0/\sqrt{\epsilon_1}$ and $Z_2 \equiv  Z_0/\sqrt{\epsilon_2}$ are wave impedances in the upper and lower semi-infinite regions with dielectric constants of $\epsilon_1$ and $\epsilon_1$, respectively.}
  \label{fig:2}
\end{figure}

The optical properties of the metal-graphene plasmonic grating in the sub-wavelength limit ($\Lambda<\lambda$) can be modeled by an equivalent two-port circuit at the junction of two semi-infinite transmission lines with impedances $Z_0/\sqrt{\epsilon_1}$ and $Z_0/\sqrt{\epsilon_2}$, that represent the upper and lower regions respectively, as shown in Fig.~\ref{fig:2} ($Z_0=377$ $\Omega $, free space impedance). The graphene can be described by a Drude conductivity
\begin{equation}\label{eq:1}
  \frac{1}{\sigma(\omega)}=\frac{(1-i\omega/\Gamma)}{\sigma_0}=R_G-i\omega L_G
\end{equation}
where $\sigma_0 \equiv ne\mu$ represents the DC sheet conductivity of a graphene layer with carrier concentration $n$ and mobility $\mu$, and $\Gamma \equiv {ev_F}/{\sqrt{\pi n}\mu\hbar}$  is the scattering rate.  From \eqref{eq:1}, the graphene may be modeled by its ohmic resistance, $R_G=\sigma_0^{-1}$, in series with its kinetic inductance, $L_G=(\sigma_0\Gamma)^{-1}$\cite{Yao2013}. $R_G$ and $L_G$ must each be multiplied by a geometrical factor of $w/\Lambda$ to account for the filling factor in this periodic structure.  The conducting contacts act as a capacitive grid \cite{Ulrich1967,Whitbourn1985} that can be described by a capacitance $C_M = 2\epsilon_0\bar\epsilon\Lambda  \ln(\csc(\pi w/2\Lambda))/\pi$, where $\bar\epsilon = (\epsilon_1 + \epsilon_2)/2$ is the average dielectric permittivity.  The finite size graphene channels contribute to an additional parallel capacitance \cite{Alu2008}, to give a total capacitance of $C = C_M + C_G = 2\epsilon_0\bar\epsilon\Lambda  \ln(2\csc(\pi w/\Lambda))/\pi$. As shown in the supplementary section (Fig. S2), this circuit accurately models the transmission, reflection, and absorption for the fundamental plasmon mode. The plasmon resonance frequency of this circuit is found to be
\begin{equation}\label{eq:2}
\omega_0^2 =\frac{e^2v_F\sqrt{\pi}}{2\hbar}\frac{\sqrt{n}}{ w\epsilon_0\bar\epsilon  \ln[2\csc(\pi w/\Lambda)]}
\end{equation}
As noted earlier, the resonant frequency scales in proportion to $n^{1/4}w^{-1/2}$, as for the case of uncontacted graphene ribbons considered in \cite{Ju2011}, indicating that $\omega_0$ can be tuned through the application of a gate voltage or by adjusting the graphene width. The resonant frequency blue-shifts weakly with increasing the duty cycle $w/\Lambda$, but in all of the cases considered here the resonance frequency is lower than that of an uncontacted graphene ribbon of the same width.

The plasmon linewidth, computed from the the equivalent circuit model is found to be
\begin{equation}\label{eq:3}
  \Delta \omega = \Gamma + \frac{\pi (Z_1^{-1} + Z_2^{-1})}{2\epsilon_0\bar\epsilon\Lambda\ln[2\csc(\pi w/\Lambda)]}
\end{equation}
The first term in \eqref{eq:3} is the conventional Drude linewidth, which is constrained by the mobility and carrier density, while the second term describes the radiative linewidth of the plasmon, which does not depend on the graphene quality or material properties. 

The equivalent circuit model can also be used to predict the condition under which maximum power is delivered to the graphene layer (Supplementary Section SS2).  The maximum on-resonant graphene absorption is achieved when the material scattering rate $\Gamma$ and radiative decay rates are equal, which also corresponds to the impedance matching between two dissimilar media\cite{Balanis2012,Alu2008}. For the parameters considered in Fig.~\ref{fig:1}a, this matching condition occurs when $\Lambda \approx 23w$, which is consistent with Fig.~\ref{fig:1}c. 

\begin{figure}
  \centering
  \includegraphics[scale=1.5]{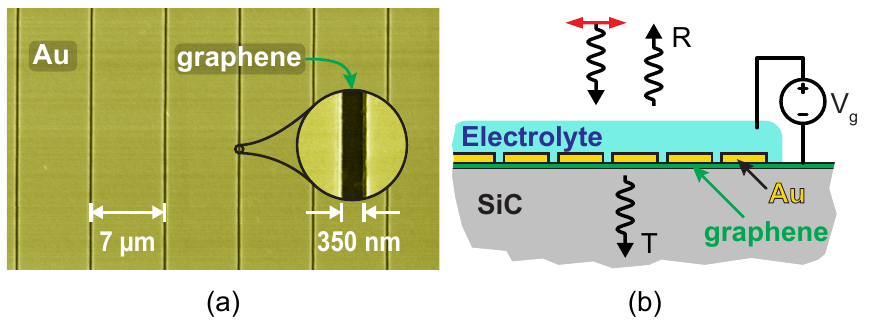}
  \caption{(a)  The false-colored SEM image of the gold-graphene grating (top view). $w$=350 nm, $\Lambda=7$ $\mu$m. (b) Diagram of the device with electrolyte top gate, and the reflection/transmission measurement scheme. The incident beam is polarized perpendicular to the gold strips.}
  \label{fig:3}
\end{figure}

\begin{figure}
  \centering
  \includegraphics{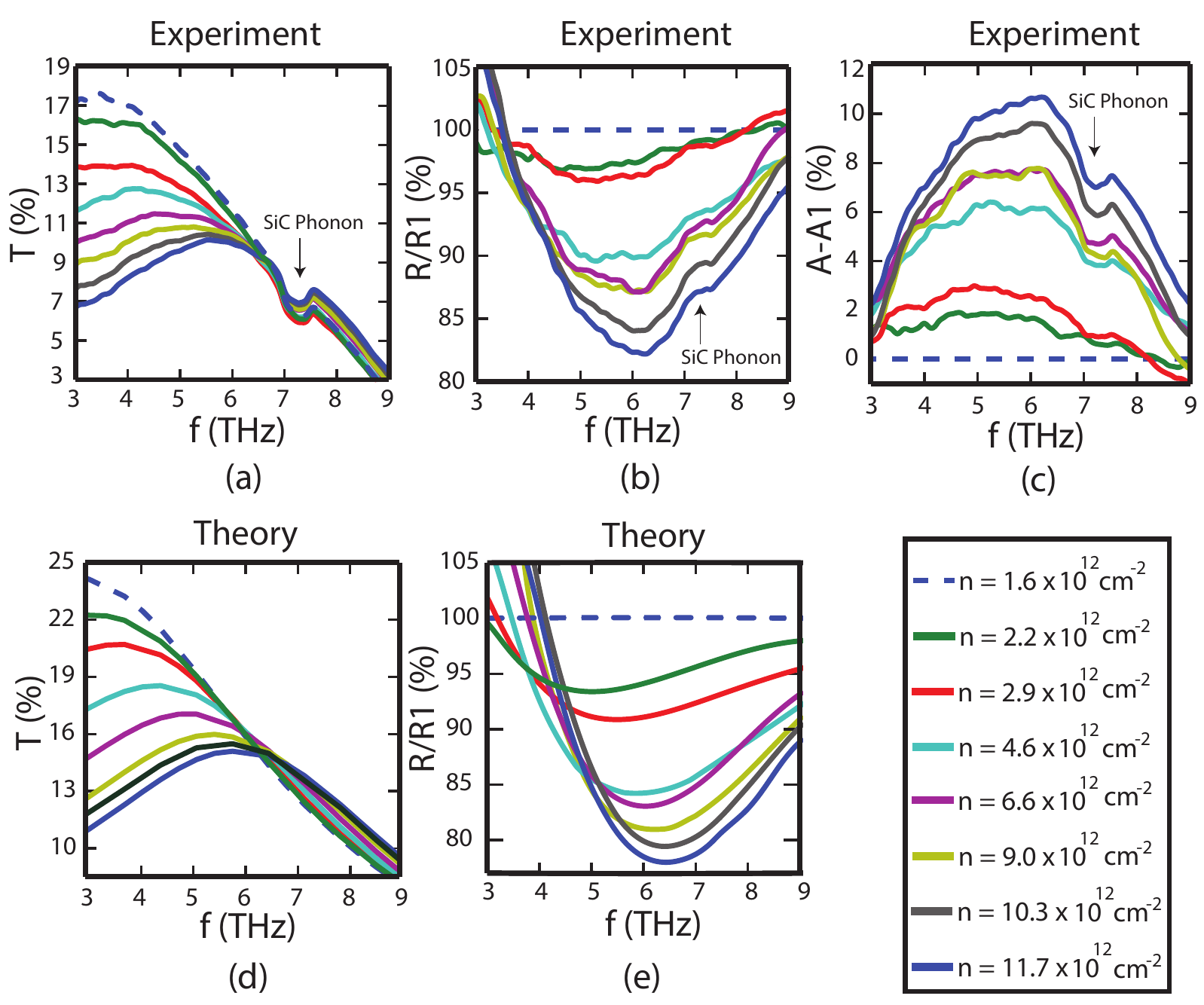}
  \caption{(a) Measured transmission ($T$) for different graphene carrier densities tuned by applying a gate voltage $V_g$. (b) Measured Reflection ($R$) off the device for different carrier densities normalized to the lowest carrier density ($n=1.6 \times 10^{12}$ cm$^{-2}$) data. (c) Control of the device absorption (relative to the lowest carrier density) by electrically tuning the graphene carrier density ($A=1-R-T$). (d) Finite element calculations of transmission for different carrier densities . (e) Finite element calculations of the normalized reflection. $w=350$ nm, $\Lambda=7$ {$\mu$}m,  $\mu=1,010$ cm$^{2}$V$^{-1}$s$^{-1}$ at $n=5 \times 10^{12}$ cm$^{-2}$. The feature at 7.2 THz is the phonon resonance of the substrate (SiC). }
  \label{fig:4}
\end{figure}

Fig.~\ref{fig:3} shows a scanning-electron micrograph image of a device with $w=350$ nm and $\Lambda=7\,\mu$m that was used to study the hybrid metal-graphene plasmons.  Fig.~\ref{fig:4}a shows the measured transmission as a function of frequency for different carrier density levels tuned by application of the gate voltage $V_g$. A resonant peak is observed in the transmission, which grows in strength and shifts to higher frequency with increasing carrier density. In reflection, the plasmon resonance exhibits a minimum that also becomes stronger and blue-shifts as the carrier density is increased (Fig.~\ref{fig:4}b). In this figure, we present the reflection normalized to the lowest carrier density data to exhibit the plasmon resonance dip more clearly. The measured absorption ($A=1-R-T$) is presented in  Fig.~\ref{fig:4}c, showing how the frequency and strength of THz resonant absorption can be controlled by tuning the carrier density with a gate voltage.

Finite element calculations of the same measured quantities presented in Fig.~\ref{fig:4}a and ~\ref{fig:4}b respectively are shown in Fig.~\ref{fig:4}d and e, showing agreement with the experimental observations. 

As with isolated graphene ribbons, the resonant frequency can also be changed by tailoring the width of the graphene channel, as predicted in \eqref{eq:2}, and confirmed experimentally in the Supplementary Section.  These results demonstrate how the hybrid metal-graphene resonances can be designed and tuned to produce strongly enhanced absorption at a chosen resonant frequency.  These hybrid plasmon modes could also be incorporated in graphene-integrated metamaterials\cite{Mousavi2013,Yao2013,Ferreira2012}, where the metal-graphene plasmon enhances the metamaterial resonance.

\begin{figure}
  \centering
  \includegraphics{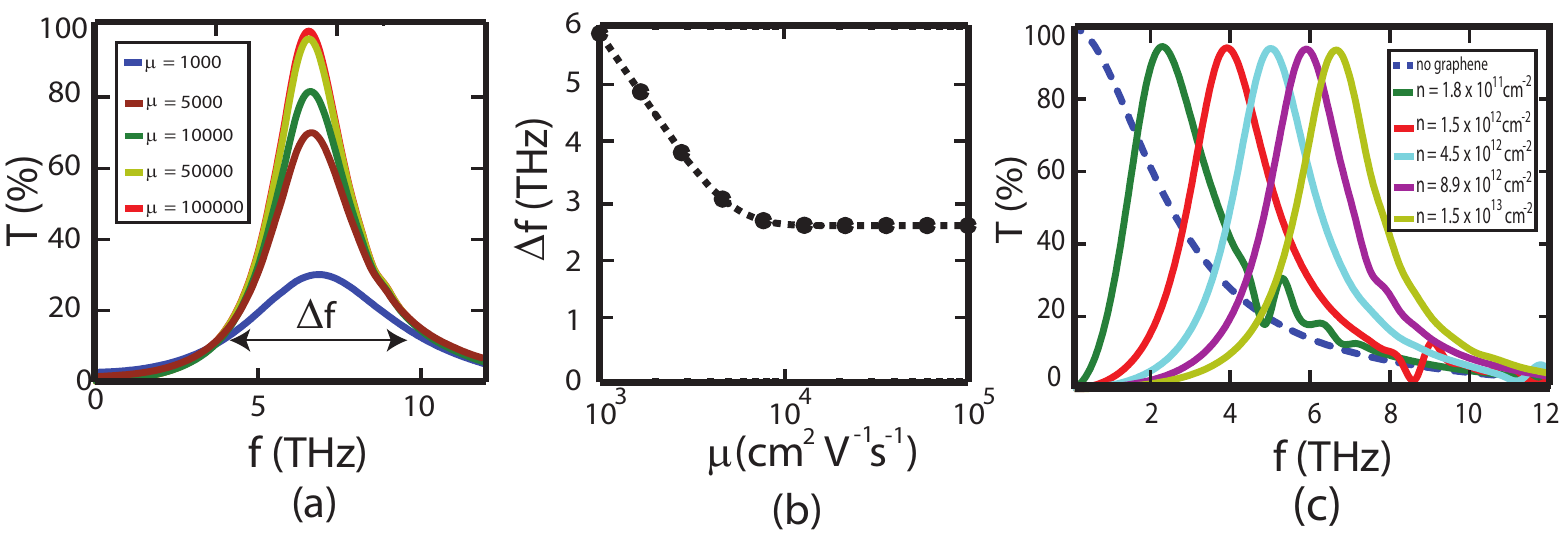}
  \caption{(a) Calculated transmission through the metal-graphene grating for different graphene mobility ($\mu$) and $n=1.5 \times 10^{13}$ cm$^{-2}$.  (b) The plasmon resonance width as a function of graphene mobility. (c) Transmission through the meta-graphene grating for different carrier density levels. $w=350$ nm, $\Lambda=7$ $\mu$m, $\mu=50,000$ cm$^2$V$^{-1}$s$^{-1}$ at $n=1.5 \times 10^{13}$ cm$^{-2}$. In all these results, the surrounding material was assumed to be uniform with $\epsilon=5$.}
  \label{fig:5}
\end{figure}

Finally, we note that these metal-graphene plasmonic structures can exhibit near 100\% resonant transmission in a high mobility graphene sample, a feature that could be very useful in THz transmission filters or modulators.  Fig.~\ref{fig:5}a shows the calculated power transmission spectrum $T(f)$ for the case of $w/\Lambda = 1/20$, and for graphene mobilities ranging from 1,000 to 100,000 cm$^2$V$^{-1}$s$^{-1}$ ($n=1.5 \times 10^{13}$ cm$^{-2}$). When the graphene mobility is increased, the graphene absorption decreases, but is replaced by a resonant peak in the transmission that approaches 100\% transmission in the limit of high mobility. Again, we note that this resonance shifts to zero frequency when the graphene is absent or charge-neutral, proving that the inductive graphene channel is essential to support the plasmonic resonance. As shown in  Fig.~\ref{fig:5}b, the spectral width of this resonance decreases inversely with the mobility, but reaches a plateau in the limit of high mobility. Above this point, the plasmon linewidth is dominated by radiation damping, and cannot be further reduced by improving the material quality, as predicted by \eqref{eq:2}. In contrast to isolated graphene ribbons, the plasmons in metal-contacted graphene are naturally radiative -- a feature that can have important consequences in tunable graphene emitters.  Fig.~\ref{fig:5}c demonstrates the tunability of the near 100\% resonant transmission through changing the graphene carrier density.  The calculated transmission spectra also reveal the presence of higher order plasmon modes that that are not described by the simple equivalent circuit model of Fig.~\ref{fig:2}.  The Supplementary Material briefly considers these higher order modes and how they can be optimized.

\section*{Methods}
\textbf{Sample and Device Preparation:}
A single layer of graphene was formed on 8mm $\times$ 8mm semi-insulating (resistivity $> 10^{10}$ $\Omega\cdot$cm) (0001)6H-SiC terraces by the Si sublimation process in an Ar ambient. The substrates, misoriented from the basal plane by approximately 0.1 deg, were etched in H2 prior to graphene synthesis \cite{Nyakiti2012}. Gold strips were fabricated on top of graphene using electron-beam lithography followed by Cr/Au thermal evaporation (Cr as the adhesion layer), and a lift-off process. The Au strips are 1.5 mm long and the whole grating is 1.5 mm wide creating a grating that has a 2.25 mm$^2$ area. To electrically isolate the grating from other parts of the graphene/SiC chip, a narrow ribbon (7 $\mu$m) was defined by electron-beam lithography using PMMA resist as a mask and oxygen plasma to remove the unmasked areas. Finally, electrolyte (Polyethylene oxide/LiClO$_4$) was drop-cast on the sample as the top gate. The gate voltage was applied between the grating device and the other electrically isolated part of the SiC graphene substrate.

\textbf{FTIR Measurement:} Far infrared simultaneous transmission/reflection measurements are performed in a BOMEM DA-8 FTIR system with mercury lamp as a source and two 4 K silicon composite bolometers as detectors. A polarizer is placed in the beam path and only passes polarization perpendicular to metal strips. The 1.5 $\times$ 1.5 mm$^2$ metal-graphene grating device is mounted on a copper plate with a 1.5 mm diameter aperture. The incident THz beam illuminated the back of the device making an angle about 10$^o$ from the normal. One bolometer is located on the transmitted beam path and one at the reflection side. A separate measurement on the sample without electrolyte was carried out to find and remove the electrolyte effect on the measured data.

\textbf{Numerical simulations:} Frequency-domain finite element calculations were performed on a unit cell of the metal-graphene grating on top of the SiC substrate (refractive index=3) with periodic boundary condition. The electrolyte on top of grating was modeled as a dielectric (refractive index=1.7). Currents, fields, and charge density in graphene and metal were calculated. Transmission and reflection of an incident plane-wave, polarized perpendicular to the metal strips, were also calculated. In the carrier-density-dependent calculations, a constant scattering rate was assumed for graphene. Mobility was taken to be 1,010 cm$^2$V$^{-1}$s$^{-1}$ at $n=5 \times 10^{12}$ cm$^{-2}$, based on van der Pauw Hall measurements taken on the full graphene on SiC sample before the processing. In the finite element calculations Fermi-level pinning at graphene-metal junction \cite{Khomyakov2010} was ignored. A constant Fermi level across the graphene channel and zero graphene-metal contact resistance were assumed. The close agreement between experimental results and theory suggest that the Fermi-level pinning and non-zero contact resistance effects are negligible in the devices we studied.  However, we expect that they should have a noticeable effect for narrow graphene channels ($<$100 nm) \cite{Khomyakov2010}.

\bibstyle{unsrt}
\bibliography{Jadidi-NatNano-2015}

\section*{Author Contributions}
M.M. Jadidi, A.B. Sushkov, H.D. Drew, M. S. Fuhrer and T.E. Murphy conceived the experiments. M.M. Jadidi conceived the theory, finite element calculations, circuit modeling and device fabrication. M.M. Jadidi and A.B. Sushkov carried out the FTIR measurements. R.L. Myers-Ward, A.K. Boyd, K.M. Daniels and D.K. Gaskill synthesized the graphene on SiC. All authors contributed to the manuscript.

\section*{Acknowledgements}
This work was sponsored by the US ONR (N000141310865) and the US NSF (ECCS 1309750). Work at NRL was supported by the Office of Naval Research.

\end{document}